**Title:** EXAMINATION OF SUPERNETS TO FACILITATE INTERNATIONAL TRADE FOR INDIAN EXPORTS TO BRAZIL*

**Authors**: Evan Winter[1], Anupam Shah[1], Ujjwal Gupta[1,5], Anshul Kumar[1], Deepayan Mohanty[2], Juan Carlos Uribe[3], Aishwary Gupta[4] and Mini P. Thomas[5]

**Affiliations:**

[1] Truffles Tech Pte Ltd

[2] former Vice-Chairman, Cargill India

[3] Head, LATAM - COFCO International

[4] DeFi Chief of Staff, Polygon Labs

[5] Department of Economics & Finance, Birla Institute of Science and Technology Pilani, Hyderabad

*Corresponding author: Ujjwal Gupta (ujjwal@truffles.one)

June 2023

# Abstract

The objective of this paper is to investigate a more efficient cross-border payment and document handling process for the export of Indian goods to Brazil.

The paper is structured into two sections:

- Explain the problems unique to the India-Brazil trade corridor by highlighting the obstacles of compliance, speed, and payments.
- Propose a digital solution for India-brazil trade utilizing Supernets, focusing on the use case of Indian exports. The solution assumes that stakeholders will be onboarded as permissioned actors (i.e. nodes) on a Polygon supernet.

By engaging trade and banking stakeholders, we ensure that the digital solution results in export benefits for Indian exporters, and a lawful channel to receive hard currency payments. The involvement of Brazilian and Indian banks ensures that Letter of Credit (LC) processing times and document handling occur at the speed of blockchain technology. The ultimate goal is to achieve faster settlement and negotiation periods while maintaining a regulatory-compliant outcome, so that the end result is faster and easier, yet otherwise identical to the real-world process in terms of export benefits and compliance.

# Keywords

Trade Finance, Letter of Credit, Cross-Border Payments, Blockchain, Export Negotiation

**India & Brazil Trade Overview:**

India and Brazil are two major economies in the world and have a long-standing trade relationship. The two countries are members of the BRICS group (along with Russia, China, and South Africa), which aims to enhance economic cooperation among its members. India and Brazil's bilateral trade has been steadily increasing over the years.
India's exports to Brazil mainly include pharmaceuticals, chemicals, textiles, auto components, and machinery, while Brazil's exports to India are mainly soybean, crude oil, sugar, and coffee.

In 2021, India exported $6.77 billion to Brazil, mainly in refined petroleum, pesticides, and packaged medicaments. Over the last 26 years, Indian exports to Brazil grew at an annualized

rate of 16%. Meanwhile, Brazil exported $4.9 billion to India in 2021, focusing on crude petroleum, soybean oil, and gold. Exports from Brazil to India increased at an annualized rate of 10.5% during the same period. This highlights the significant growth of trade between the two nations over the past two decades.

| India Exports to Brazil | Value | Year | Brazil Exports to India | Value | Year |
|---|---|---|---|---|---|
| Miscellaneous chemical products | $1.20B | 2021 | Animal, vegetable fats and oils, cleavage products | $2.37B | 2021 |
| Mineral fuels, oils, distillation products | $1.12B | 2021 | Mineral fuels, oils, distillation products | $1.85B | 2021 |
| Organic chemicals | $644.99M | 2021 | Pearls, precious stones, metals, coins | $813.26M | 2021 |
| Machinery, nuclear reactors, boilers | $475.89M | 2021 | Sugars and sugar confectionery | $228.52M | 2021 |
| Vehicles other than railway, tramway | $377.39M | 2021 | Ores slag and ash | $154.17M | 2021 |
| Pharmaceutical products | $375.41M | 2021 | Machinery, nuclear reactors, boilers | $123.49M | 2021 |
| Aluminum | $225.86M | 2021 | Iron and steel | $105.92M | 2021 |
| Man Made filaments | $215.62M | 2021 | Salt, sulfur, earth, stone, plaster, lime and cement | $88.60M | 2021 |
| Rubbers | $194.73M | 2021 | Miscellaneous chemical products | $65.84M | 2021 |
| Iron and steel | $193.22M | 2021 | Cotton | $65.77M | 2021 |
| Tanning, dyeing extracts, tannins, derivatives, pigments | $144.86M | 2021 | Organic chemicals | $60.13M | 2021 |
| Electrical, electronic equipment | $115.66M | 2021 | Edible vegetables and certain roots and tubers | $54.88M | 2021 |
| Articles of iron or steel | $112.32M | 2021 | Wood and articles of wood, wood charcoal | $49.89M | 2021 |
| Plastics | $102.70M | 2021 | Electrical, electronic equipment | $29.36M | 2021 |
| Man Made staple fibers | $90.44M | 2021 | Pharmaceutical products | $23.61M | 2021 |
| Ores slag and ash | $81.08M | 2021 | Essential oils, perfumes, cosmetics, toiletries | $22.37M | 2021 |
| Optical, photo, technical, medical apparatus | $76.99M | 2021 | Coffee, tea, mate and spices | $21.08M | 2021 |
| Inorganic chemicals, precious metal compound, isotope | $46.18M | 2021 | Paper and paperboard, articles of pulp, paper and board | $16.87M | 2021 |
| Ceramic products | $43.87M | 2021 | Beverages, spirits and vinegar | $15.96M | 2021 |
| Articles of apparel, not knit or crocheted | $41.58M | 2021 | Oil seed, oleaginous fruits, grain, seed, fruits | $13.86M | 2021 |
| Glass and glassware | $34.17M | 2021 | Residues, wastes of food industry, animal fodder | $13.72M | 2021 |
| Aircraft, spacecraft | $33.31M | 2021 | Rubbers | $13.03M | 2021 |
| Cotton | $30.18M | 2021 | Optical, photo, technical, medical apparatus | $12.94M | 2021 |
| Tools, implements, cutlery of base metal | $22.92M | 2021 | Raw hides and skins (other than furskins) and leather | $12.20M | 2021 |
| Essential oils, perfumes, cosmetics, toiletries | $21.17M | 2021 | Vehicles other than railway, tramway | $11.99M | 2021 |
| Soaps, lubricants, waxes, candles, modeling pastes | $20.84M | 2021 | Inorganic chemicals, precious metal compound, isotope | $10.45M | 2021 |

## Literature Review

This literature review examines existing research on cross-border trade between India and Brazil, focusing on the challenges and obstacles faced in terms of compliance, speed, and payments. We also explore the potential of blockchain technology to address these challenges, improve efficiency, and streamline export processes. Additionally, we present the theoretical foundation for our proposed digital solution that utilizes Polygon's Supernets technology to enhance the India-Brazil trade corridor.

**Challenges in India-Brazil Cross-Border Trade**

Literature highlights various challenges in the bilateral trade relationship between India and Brazil, including high transaction costs, lack of transparency, and inefficiencies in existing trade and payment systems (Smith et al., 2018; Gupta & Kumar, 2020). Researchers have also emphasized the impact of regulatory compliance and the protracted process of obtaining Letters of Credit (LC) on trade efficiency (Gupta & Kumar, 2020; Shrivastava, 2019).

Niepmann and Schmidt-Eisenlohr (2017) investigated the extent to which firms utilized special trade finance products offered by banks, finding that 15% of global exports were settled by letters of credit and documentary collections, with letters of credit predominantly used for exports to countries with intermediate contract enforcement.

Chaturvedi (2011) analyzed South-South cooperation in health and pharmaceuticals, specifically the bilateral partnership between India and Brazil. The study revealed that Brazil was a top export destination for Indian pharmaceutical products, with the pharmaceutical sector being a significant source of Indian investment in Brazil.

Mishra et al. (2015) estimated a gravity model of international trade between India and other BRICS countries, discovering that per capita GDP and transport costs significantly impacted India's trade with Brazil, Russia, China, and South Africa. However, exchange rate, inflation, and import-to-GDP ratio were found to have an insignificant impact on India's bilateral trade with other BRICS countries.

**Blockchain Technology and International Trade**

Blockchain technology has emerged as a potential solution to numerous challenges faced in international trade. Tapscott and Tapscott (2016) and Swan (2015) emphasize the potential of blockchain technology to revolutionize trade processes by providing a secure, transparent, and efficient platform for cross-border transactions. Treleaven et al. (2017) and Saberi et al. (2019) focus specifically on the role of blockchain in streamlining trade finance, LC processing, and document handling.

**Polygon Supernet and Permissioned Nodes**

The concept of supernets and permissioned nodes has garnered attention as a means of improving the security and efficiency of blockchain networks. Buterin et al. (2020) and Miers et al. (2013) discuss the use of supernets and permission nodes in blockchain networks and their potential to enhance trade processes. Our proposed digital solution builds on this foundation by using Polygon supernet technology to onboard trade and banking stakeholders as permissioned nodes, streamlining the India-Brazil trade process.

In summary, this literature review covers the challenges faced in cross-border trade between India and Brazil, the potential of blockchain technology to address these challenges, and the theoretical underpinnings of our proposed digital solution. By incorporating these findings, our study aims to develop a well-informed digital solution leveraging blockchain technology to improve cross-border payment and document handling processes in the India-Brazil trade corridor.

## Methodology

**Role of Polygon Supernet Technology:**

In our methodology, we also consider the role of our proposed technology solution. This solution utilizes the Polygon supernet, a type of blockchain network, to enhance the trade processes between India and Brazil. The technology enables secure, transparent, and efficient cross-border transactions by allowing trade and banking stakeholders to become permissioned nodes. This system could significantly streamline document handling and cross-border payment processes, thus promoting trade efficiency in the India-Brazil corridor.

The key step in our methodology revolves around the development of a Polygon Supernet-based prototype to represent the trade process between India and Brazil. This would include a network of permissioned nodes that signify crucial stakeholders in the trading process, such as banks and customs authorities. Each node within the network would have distinct roles and permissions, thereby ensuring a secure and appropriate distribution of information.

In this system, the issuance of Letters of Credit (LC) and the documentary process take place via the minting of a non-fungible token (NFT) from the infrastructure smart contract.The NFT represents the unique details of each trade transaction, invoicing including shipment details, financial documents, and LCs.

Contrary to typical public blockchain networks, in our proposed system, the banks or financial institutions act as the sole validators. This ensures that the network is secure and that the data integrity is maintained since these institutions are trusted parties in the international trade finance process.

**Document verification and amendment:**

Merkle proofs play a crucial role. Merkle proofs are cryptographic constructs that allow the tracing of all document histories, enabling a clear visibility into the entire document's changes over time. They allow participants in the trade process to keep track of every change, ensuring transparency and accountability in trade transactions.

Furthermore, we address the high-level privacy concerns often associated with trade documents, particularly with respect to final document iterations after several amendments among banks. We propose the use of zero-knowledge proofs, a cryptographic technique which allows an authorized party to prove knowledge of specific information without revealing that information. This method enables secure verification of the authenticity of a trade document without exposing the document's details or necessitating a cross-reference with source documents also leveraging ERC-2535, also known as Diamond Standard, to construct on-chain business logic with complete smart contract upgradability. This allows us to create more complex systems and contracts that can evolve over time without losing their state or transaction history.

Also, The methodology of this study involves using three key indices to analyze the trade relationship between India and Brazil. These indices helps us for understanding the improvements in trade volume, competitive advantage, and value addition brought by the Polygon Supernet technology in the India-Brazil trade corridor.

- Trade Intensity Index (TII),
- Revealed Comparative Advantage (RCA) Index, and
- Trade in Value Added (TiVA)

These indices provide insights into the current state of trade between the two countries and help identify areas for potential improvement. The paper proposes a Polygon supernet-based digital solution to enhance trade efficiency between India and Brazil by streamlining cross-border payment and document handling processes.

**Trade Intensity Index (TII):**

The TII measures the intensity of trade between two countries relative to their overall trade volumes. It is calculated using the formula:

TII = (Xij / Xi) / (Mwj / Mw)

Where:
Xij is the export value of country i (India) to country j (Brazil)
Xi is the total export value of country i (India) to the world
Mwj is the total world exports to country j (Brazil)
Mw is the total world exports

**Revealed Comparative Advantage (RCA) Index:**

The RCA Index assesses the relative advantage of a country in producing and exporting a particular product. It compares a country's share of a product's global exports with its share of total global exports. An RCA greater than one indicates a comparative advantage, while an RCA less than one signals a comparative disadvantage.

$$\frac{\sum_d x_{isd} / \sum_d X_{sd}}{\sum_{wd} x_{iwd} / \sum_{wd} X_{wd}}$$

where s is the country of interest, d and w are the set of all countries in the world, i is the sector of interest, x is the commodity export flow and X is the total export flow.

**Trade in Value Added (TiVA):**

TiVA measures the value added by a country in producing and exporting a particular product, accounting for the contributions of all countries involved in the production process.

# Data Analysis

**Trade Intensity Index (TII):**

For India and Brazil, the TII was calculated using 2021 trade data:

India's exports to Brazil (2021): $6.77 billion

India's total exports to the world (2021): $403 billion

World exports to Brazil (2021): $225 billion

Total world exports (2021): $21 trillion

TII (India to Brazil) = ($6.77 billion / $403 billion) / ($225 billion / $21 trillion) = 1.567

A Trade Intensity Index (TII) of 1.567 between India and Brazil indicates that the intensity of trade between these two countries is greater than what would be expected given their share in world trade.
More specifically, it means that the proportion of India's exports going to Brazil is significantly larger than the proportion of the world's exports that go to Brazil.
In other words, India trades more intensively with Brazil than the average country does, indicating a stronger trade relationship between these two countries.

**Revealed Comparative Advantage (RCA) Index:**

For the product category of pharmaceuticals:

India's global exports of pharmaceuticals (2021): $21.7 billion
India's total global exports (2021): $403 billion
India's share of pharmaceutical exports to total exports: 21.7/403

Brazil's global exports of pharmaceuticals (2021): $1.36 billion
Brazil's total global exports (2021): $288 billion
Brazil's share of pharmaceutical exports to total exports: 1.36/288

Global exports of pharmaceuticals (2021): $806 billion
Total global exports (2021): $21 trillion
Global share of pharmaceutical exports to total exports: 0.806/21

RCA (India, Pharmaceuticals) = 1.40

RCA (Brazil, Pharmaceuticals) = 0.12

The RCA of 1.40 indicates that India has a significant comparative advantage in the production and export of pharmaceuticals compared to Brazil which has an RCA of 0.12.

**Trade in Value Added (TiVA):**

For the product category of automotive parts:

According to the Organisation for Economic Co-operation and Development (OECD), in 2016, India's value added in the production of automotive parts was $10.2 billion, while Brazil's value added was $6.4 billion. We can calculate the TiVA percentage for both countries as follows:

India's TiVA percentage: ($10.2 billion / ($10.2 billion + $6.4 billion)) * 100 = 61.4%
Brazil's TiVA percentage: ($6.4 billion / ($10.2 billion + $6.4 billion)) * 100 = 38.6%

The TiVA percentages of 61.4% for India and 38.6% for Brazil indicate that India contributes more to the value added in the production of automotive parts compared to Brazil.

The proposed digital solution could potentially improve the TIVA of India and Brazil in several ways:

1. Reducing transaction costs and time associated with the trade process, leading to increased trade: By providing a more efficient and secure cross-border payment and document handling process, the proposed digital solution can help reduce the costs and time associated with the trade process. This can potentially lead to an increase in trade between India and Brazil, which can improve the TIVA of both countries.

2. Increasing trust between Indian and Brazilian banks, leading to increased trade in value-added products where India and Brazil have a comparative advantage: By providing a secure and efficient cross-border payment process, the proposed solution can potentially increase the trust between Indian and Brazilian banks. This can lead to an increase in trade in value-added products, where India and Brazil have a comparative advantage. This can improve the TIVA of both countries in these products.

3. Reducing the risk associated with trade finance by leveraging the speed and security of blockchain technology, encouraging banks to finance more trade between the two countries: By providing a secure and efficient cross-border payment process, the proposed solution can potentially reduce the risk associated with trade finance. This can encourage banks to finance more trade between India and Brazil, which can improve the TIVA of both countries.

In summary, the data analysis using the TII, RCA, and TiVA indices reveals that there is room for improvement in the trade relationship between India and Brazil. The proposed digital solution, which leverages blockchain technology for efficient cross-border payment and document handling processes, could potentially enhance trade efficiency and boost the TII, RCA, and TiVA of both countries.

## Problem Statement

The main reason why trade between India and Brazil is challenging is that both countries have strict capital controls that impact trade. In Brazil and India, hard currencies ("USD / EUR etc.")can only be paid through banks. Corporate treasuries do not have exposure to foreign currency and only banks can provide the service of converting local currency (INR and BRL) into USD etc.

This makes it difficult for Brazilian importers to pay for Indian exports, because

1) Brazilian importers must balance their hard currency payments – i.e. inflow of USD and outflow of USD.

2) The second barrier from the Brazilian perspective is the high cost associated with LC issuance as well as the lack of financing options for importers who do not have hard currency receivables. This high cost is coupled with manual bank processes, which render the current solution costly and time-consuming.

**Challenges & Difficulties in Using LC:**

While Letters of Credit (LCs) are a commonly used payment method in international trade, they can also present several challenges and difficulties.

1. Cost: LCs can be expensive to obtain and maintain. Banks typically charge fees for issuing and amending LCs, and these fees can add up quickly.
2. Complexity: LCs can be complex and difficult to understand, especially for businesses that are new to international trade. The rules and procedures governing LCs can be highly technical and require a lot of paperwork and documentation.
3. Disputes: Disputes can arise over the terms and conditions of LCs, especially if the documents presented do not match the requirements of the LC. Disputes can result in delays in payment or even legal action.
4. Fraud: LC fraud is a common problem in international trade. Fraudulent LCs can result in significant financial losses for the parties involved, and can damage business relationships and reputations.
5. Time-consuming: The process of obtaining and using LCs can be time-consuming, and can result in delays in the shipment of goods or in payment.

**Here is a case study regarding LC fraud that we are proposing a solution to in the paper:**

**Case Study: The Nirav Modi LC Fraud Case**

Introduction:

The Nirav Modi case is one of the largest LC fraud cases in India, involving fraudulent issuance of Letters of Undertaking (LoUs) by the Punjab National Bank (PNB) to Nirav Modi's companies. The case highlights the importance of due diligence and risk management in international trade, and the need for stronger oversight and regulation in the banking sector.

Background:

Nirav Modi, a billionaire diamond merchant, allegedly colluded with PNB officials to obtain fraudulent LoUs, which were issued without proper collateral or documentation. These LoUs were then used to obtain credit from overseas branches of Indian banks. The funds obtained through these fraudulent transactions were allegedly laundered through a complex web of shell companies and offshore accounts.

Timeline:

The fraud came to light in January 2018, when PNB filed a complaint with the Central Bureau of Investigation (CBI) alleging fraudulent transactions worth over $2 billion. The CBI subsequently filed a case against Nirav Modi, his uncle Mehul Choksi, and others for cheating and criminal conspiracy. Nirav Modi fled the country in early 2018 and is currently a fugitive from justice.

Impact:

The Nirav Modi case has had a significant impact on the Indian banking sector, with several banks reporting losses due to exposure to fraudulent transactions. The case has also highlighted the need for stronger risk management measures and oversight in the banking sector. The Indian government has taken steps to extradite Nirav Modi from the UK, where he is currently residing.

Conclusion:

The Nirav Modi case serves as a reminder that LC fraud is a global problem that can occur in any international trade, regardless of the countries involved. Businesses need to be aware of the risks of LC fraud and hence as our solution with this paper helps in solving this if PNB and its overseas partners are on the same blockchain network, they will be able to verify if duplicate LCs were issued to multiple banks and hence could prevent the fraud. This case also highlights the need for stronger regulation and oversight in the banking sector which can be done with robustness if we shift from old LC methods to modern digital LC secured through Blockchain.

## Here are the challenges from India's side:

An Indian exporter can only receive a foreign currency payment once they meet several criteria that allow banks to credit the foreign currency. Here are the points that impact the Exporter's ability to receive a USD credit under documentary acceptance or LC negotiation:

> a. Indian exporters avail export benefits and therefore require a bank realization certificate ("BRC"), which can only be issued by an Indian bank.

b. The Guaranteed Remittance ("GR") form – this is a document required for a hard currency payment against export. The exporter collects the GR form from customs and submits the form to its bank within a certain time period of receiving a foreign currency payment.

c. The Indian bank is also the primary actor that grants trade finance to the exporter and therefore in most cases, the exporter is bound to negotiate the export sale through the Indian bank. This means that the export LC negotiation or presentation of export docs under Documentary Acceptance ("DA") involves the Indian bank counters to negotiate the sale; in the current banking paradigm, this process is slow and largely manual.

Here is a depiction of the current trade process:

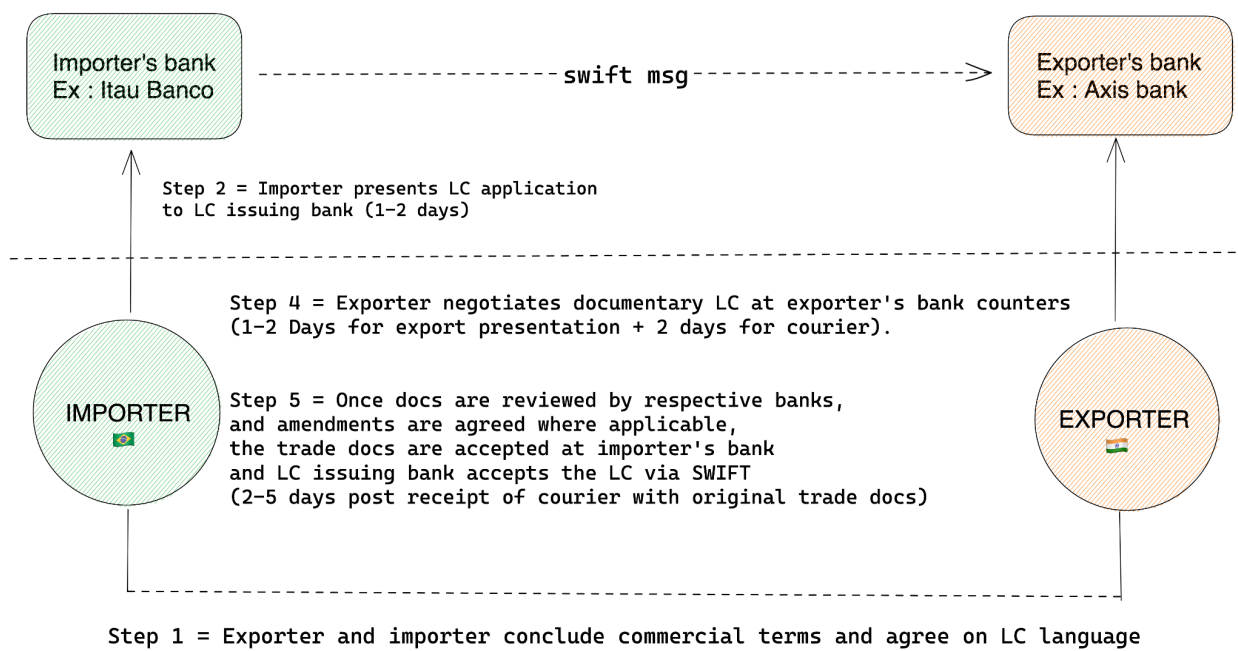

In summary, the Indian bank stakeholder is the primary actor to ensure that the exporter has complied with export regulations to receive the BRC and to whom exporters submit the GR form, thus we include the Indian bank as a stakeholder in the supernets framework, which improves the export negotiation process and streamlines the sharing of documentation. The same approach improves the DA presentation process for which the Indian bank is also the primary stakeholder.

## Solution :

As an alternative to SWIFT, we propose a blockchain-based solution, which is effective to establish security, consensus, and a digital LC trade negotiation and payment system. As more banks participate in the framework, the network effect increases:

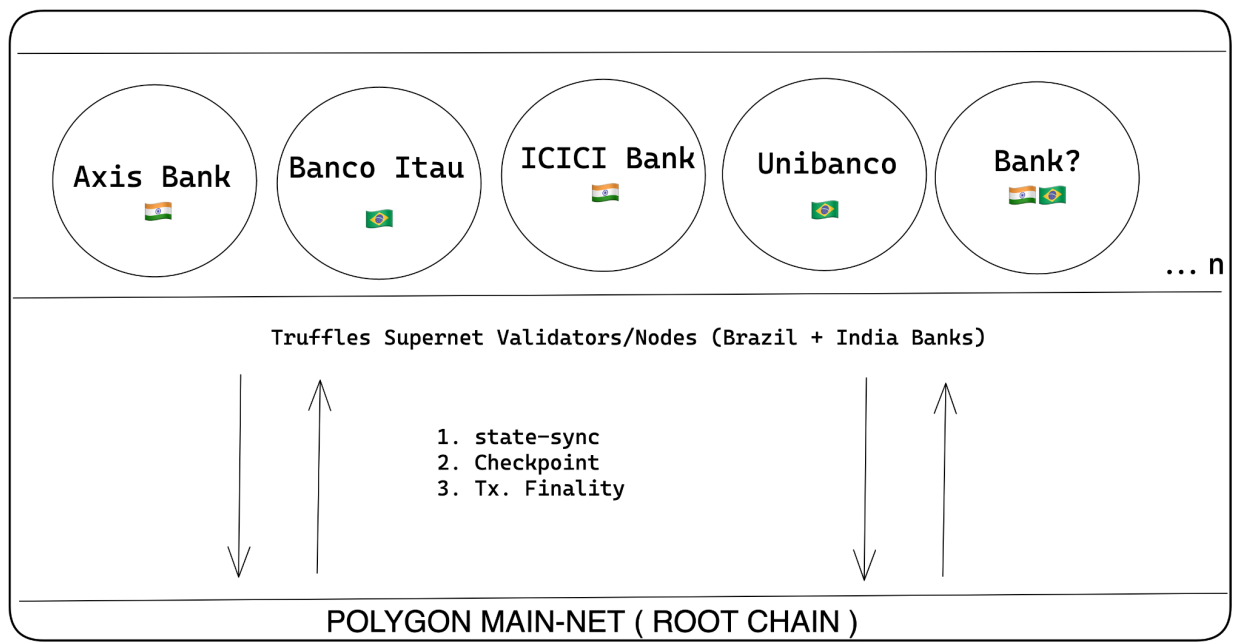

In order to participate in the network as a bank, it needs to run a supernet validator which is constantly updating the distributed ledger and securing all on-chain data via cryptography. Nodes/validators also secure the network directly by allocating computational resources which in turn grows directly as more validators join this permissioned blockchain environment referred to as supernet.

Consensus among validators is required to allow joining/removing a validator through a voting mechanism - where the network participants decide on certain decisions which are otherwise centralized in nature due to legacy systems being isolated from each other, instead of being in sync with each other as visible in the diagram above.

Here is a diagram that explains the privacy and decentralized data storage implementation:

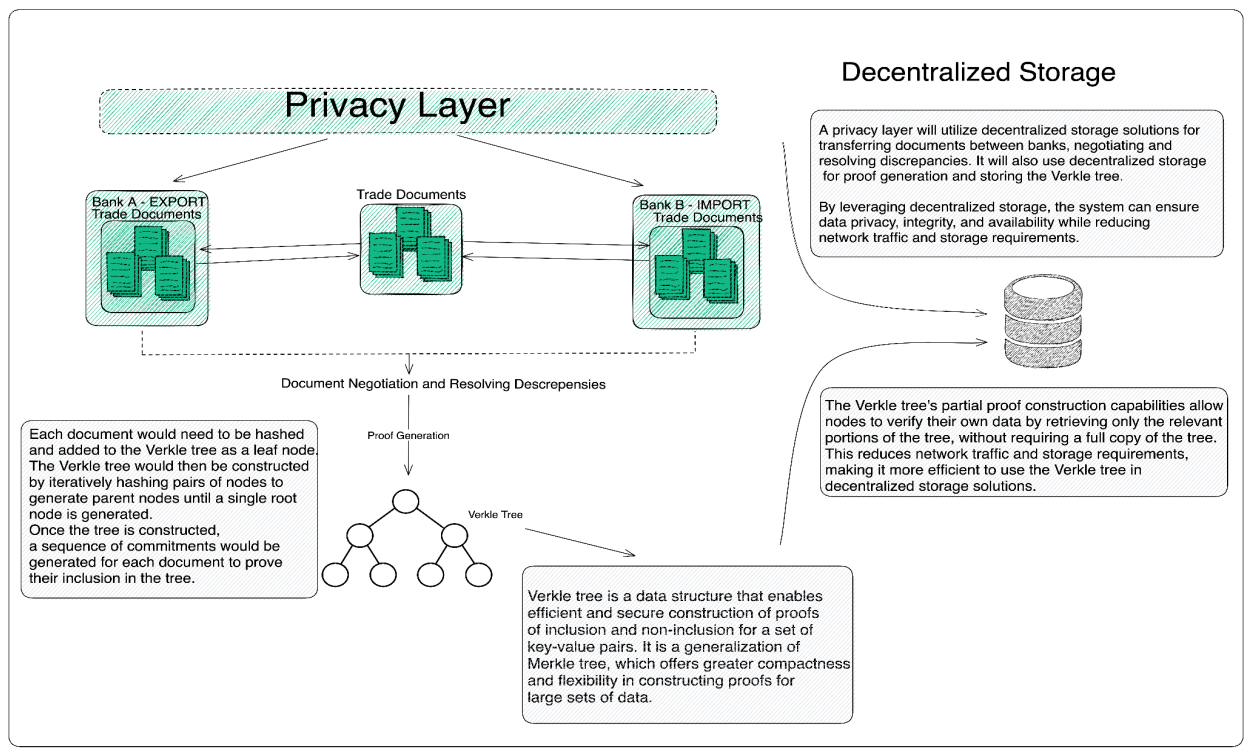

The solution is scalable and mitigates the onerous time cost in the current trade process while maintaining security and bank control over critical functions. Through the solution, banks can communicate more effectively, and reach consensus on trade docs, while not suffering trade-offs in terms of security. Payments can also be directly implemented into the blockchain solution.

Verification and amendments of documents can be traced back to all of their history using Verkle Proofs.

Certain core components that involve a very high level of privacy such as the finality of a trade document after numerous back-and-forth amendments among banks could also be secured by leveraging zero-knowledge proofs, which require an authorized party to decrypt the information by generating cryptographic proofs instead of viewing an entire document & cross-referencing with source documents.

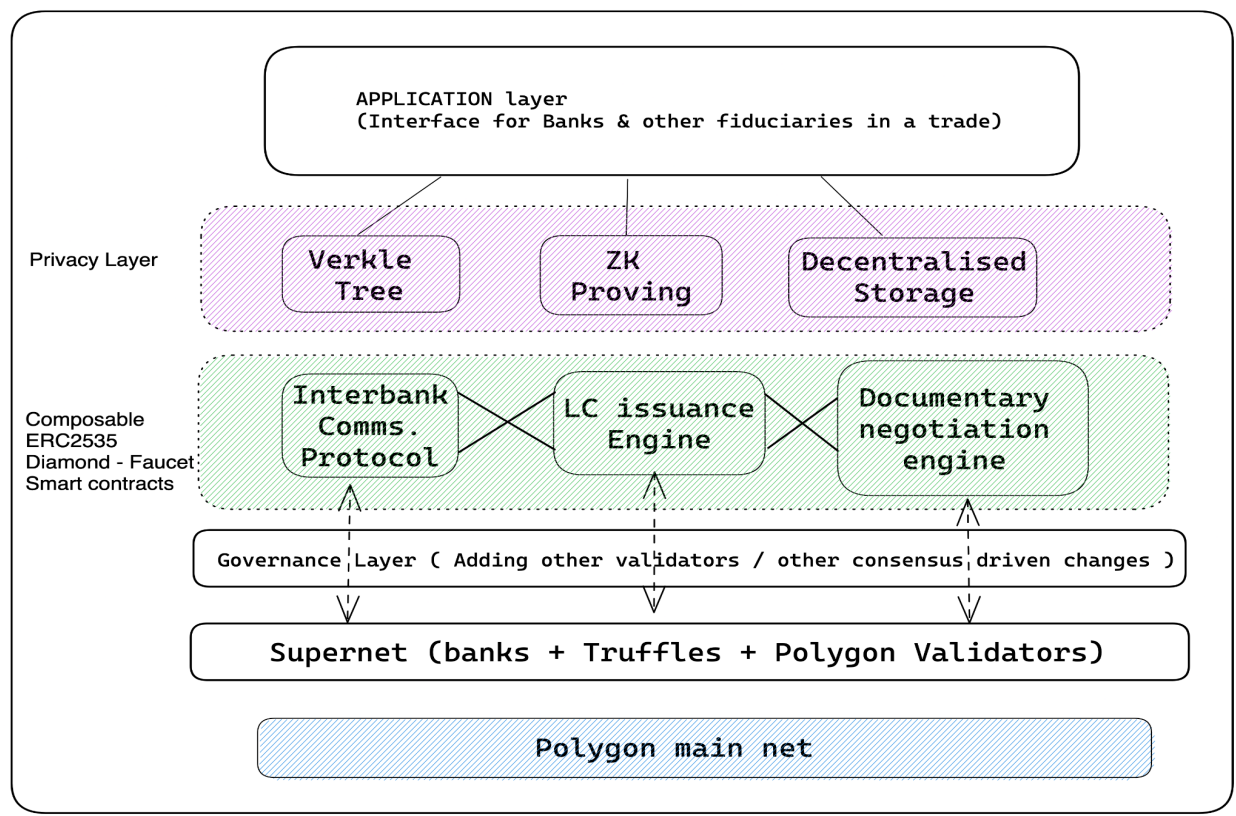

The Banking Supernet depicted above will be an app-specific child chain built on top of Polygon Mainnet (Root Chain).

As a private, firewalled blockchain built to accommodate Brazilian & Indian Banks, each bank runs the network validators/nodes, thereby decentralizing the network & creating an on-chain governance mechanism.

Our proposed solution can directly scale and build more value as other banks join the Supernet as validators, thereby providing the computational resources to scale the network. ERC2535 Contracts have the capacity to build on-chain business logic with complete smart contract upgradability.

Communication between the bank validators and document verification can now be implemented on-chain with the benefits of a 'state replicated machine'. The blockchain can be leveraged to guarantee a single source of truth for all information with complete privacy protection & a fair consensus mechanism.

All aspects of the export trade become digital, with encryption and security at the forefront. Digital documentation can be hashed to prevent tampering. The following flow can be implemented via this solution architecture with a user interface built as an application that integrates with the supernet:

1. The 4 parties to the trade are permissioned on the private chain with an on-chain identity and with roles and responsibilities.

2. Importers and exporters agree on trade terms – this can be digital or in the real world.

3. Importers, exporters and banks agree on LC language. Rather than via SWIFT, the LC is issued digitally with the metadata minted onto an NFT such that the readable PDF file contains all standard fields to an LC.

4. The NFT LC is submitted from the importer's bank to the exporter's bank and the image is read to understand the terms of the LC.

5. The exporter uploads the docs to the user interface created on the application layer for the purpose of facilitating doc uploading and transfer.

6. A trusted copy of the title docs is generated digitally by the Indian bank and measured against the token bearing the LC terms. Any communication on discrepancies is navigated through on-chain communication between the importers' and exporters' banks. All communication fees and presentation charges are expressed as gas (that can be supported from the main net contract for a smooth user experience). Therefore the costs are significantly reduced vs. manual processes. Some manual charges will remain as bank operations will still need to measure the trade docs against the LC terms and the original title docs will still need to be couriered from the exporter's bank to the importer's bank for customs clearance.

7. Once the exporter has presented docs that the LC advising bank determines are compliant with the LC terms, the documentation is encrypted and transferred on-chain to the importer's bank. The importer's bank can decrypt and measure the docs against the LC terms again using the blockchain for communication.

8.  Once the LC docs are accepted, the acceptance is brought on-chain as a meaningful event and now we can proceed to facilitate payment between the bank nodes.
    We will use onshore onramps in Brazil ideally where local currency can be on-ramped into USDC or other polygon-supported stable tokens. This replaces the process of buying foreign currency, and the Brazilian bank rather buys USDC with the destination on-chain as its self-custodial wallet, or there can be automatic facilitation of payment where the Financial Institution on-ramp partner is purely an intermediary that immediately transfers to the Indian bank wallet.

    The stable token is off-ramped and local currency at current exchange rates is credited to the exporter's bank where charges can be adjusted and then the exporter's account is credited in INR.

## Alternative Settlement Layer to USD/USDC:

While re-imagining the LC issuance and negotiation process, it is important to consider alternative settlement options that may be more beneficial to Emerging Market economies than using USD or pegged stable tokens to USD.

Although a specific solution is outside the scope of this paper, it is worth considering several high-level alternatives to USD settlement for foreign trade.

USD as a settlement medium is beneficial for exporters because they can reasonably expect that fiat value will be maintained at the time of shipment. For an exporter to ship, it is important that the settlement currency retains relative value. As USD dominates global commodity markets as the indexed point of value, an exporter can reasonably expect to receive the value that is preserved over long manufacturing/shipping periods. As most exporters need to monetize exports in order to purchase further inputs for a manufacturing process, the stable nature of USD is extremely important to ensure that an exporter's business margin is somewhat stable and that the payment for export will carry value over to the exporter's purchase of input goods. This is especially true for the export of value-added goods that depend on commodity inputs.

With this context, a proposed alternative to the USD as settlement currency must ensure that iterative renditions of settlement currencies preserve value and are ideally linked to future input purchases so that manufacturers continue producing, selling, and accepting value, then repeat the cycle.

**Alternatives to USD settlements:**

1. Stable token backed by a customizable basket of goods where the settlement is abstracted away from the two emerging market currencies related to the trade and the value of this settlement medium is directly linked to the export value of goods and services, while also considering input values as part of the product basket.

In other words, a stable token that is backed by a basket of value-added goods and commodity inputs that are customizable basis the particulars of a given trade corridor. The exporter will receive the value that is preserved or correlated at least somewhat with its input costs. Importers will be exposed to price fluctuations in whatever goods they are buying, but this can be mitigated by adding goods to the basket that reflect the exports from the buyer's country.

To add an example here: India exports machines to Brazil that enhance the yield of soybean crops. If the value of an LC was related to both the changing market price of the machines as well as the soybeans, then both importer and exporter may enjoy a settlement value that is correlated to the overall market performance of their industry.

**2. Settlement in INR/Reais**

This solution requires a thorough analysis of trade relationships and may involve deal-making and escrow mechanisms that ensure a payment medium can carry forward value that may be used to purchase goods in the future.

To explain further, an LC can be denominated in Reais, if a reserve of Reais can be established to provide seamless FX for the purchase of goods in the reverse direction (Ie reais are converted to INR to satisfy the exporter). The Reais can be escrowed to facilitate the purchase of Brazilian exports. Although this is a crude demonstration of this solution, it is worth noting and closely follows the actions of China, Russia, UAE, and Saudi Arabia who are engaged in trade discussions around alternative settlement functions. India has also established escrow mechanisms in the past with trade partners such as Israel and Russia.

**Annexure 1:**

Current trade processes explained (DA & LC trade processes):

Here is how a DA (Documentary Acceptance) trade works:

The comparison of Documentary Acceptance (DA) trade and Letter of Credit (LC) trade processes highlights the differences in time and costs associated with each method. The DA trade process is generally faster but may have higher costs, while the LC trade process offers more security and certainty of payment, but with longer processing times and additional costs related to credit.

**Documentary Acceptance (DA) Trade:**
- Faster process: Total time of 5-7 days, with possible delays up to 2 weeks
- Higher costs: approximately 0.5% or more for smaller trades
- Exporter maintains control of essential title documents for customs clearance
- Currency transfer through SWIFT

**Letter of Credit (LC) Trade:**
- Offers more security and certainty of payment
- Longer processing times: 1-2 weeks for presentation and negotiation
- Additional costs related to credit: depends on factors such as exporter's decision to discount the LC, strength of the LC issuance bank, and creditworthiness of parties involved
- Banks scrutinize export documents against LC terms

In summary, the choice between DA trade and LC trade depends on the priorities of the parties involved in the transaction, such as speed, costs, and security. While DA trade is generally faster, it may come with higher costs, whereas LC trade provides more security and certainty of payment at the expense of longer processing times and additional costs associated with credit.

**References:**

- Apte, S., Petrov, M., & Singh, M. (2018). Blockchain Technology: Transforming International Trade. Journal of Internet Banking and Commerce, 23(3), 1-13.
- Baumann, R., Ribeiro, F. J., Carneiro, F. L., & Araújo, M. de A. (2021). Brazil and India: Peculiar relationship with big potential (Discussion Paper No. 1). Institute for Applied Economic Research (Ipea).
- Casey, M. J., & Vigna, P. (2018). The Truth Machine: The Blockchain and the Future of Everything. St. Martin's Press.
- Chaturvedi, S. (2011). South-South Cooperation in Health and Pharmaceuticals: Emerging Trends in India-Brazil Collaborations. RIS Discussion Paper No. 172, Research and Information System for Developing Countries, Government of India, New Delhi.
- Chen, X., & Chen, C. (2021). Cross-border Payments Based on Blockchain Technology: A Review. China Economic Journal, 14(1), 1-20.
- Embassy of India. (2018). India-Brazil Bilateral Trade Relations.
- Export-Import Bank of India. (2016). Intra-BRICS Trade: An Indian Perspective (Working Paper No. 56).
- Hoekstra, M., & Hermes, N. (2018). Trade Finance in Developing Countries: A Qualitative Analysis. The European Journal of Development Research, 30(4), 670-695.
- Kamble, S. S., Gunasekaran, A., & Sharma, R. (2019). Modeling the blockchain enabled traceability in agriculture supply chain. International Journal of Information Management, 49, 263-277.
- Mishra, A. K., Gadhia, J. N., Kubendran, N., & Sahoo, M. (2015). Trade Flows between India and Other BRICS Countries: An Empirical Analysis Using Gravity Model. Global Business Review, 16(1), 107–122. https://doi.org/10.1177/0972150914553523
- Mougayar, W. (2016). The business blockchain: Promise, practice, and application of the next internet technology. John WSupeiley & Sons.
- Niepmann, F. & Schmidt-Eisenlohr, T. (2017). International trade, risk and the role of banks. Journal of International Economics, 107, 111-126, https://doi.org/10.1016/j.jinteco.2017.03.007

arXiv PREPRINT- OEC World. (2022). India-Brazil Bilateral Trade Data.
- OECD. (2018). Trade in Value Added (TiVA): Brazil Country Note. Retrieved from https://www.oecd.org/sti/ind/tiva/CN_BRA.pdf
- Polygon Team. (2023). Polygon Supernets Documentation.
- Rejeb, A., Keogh, J. G., & Treiblmaier, H. (2019). Leveraging the Internet of Things and Blockchain Technology in Supply Chain Management. Future Internet, 11(7), 161.
- Sen, S. (2017). BRICS and the new financial architecture. Economic and Political Weekly, 52(11), 25-27. https://www.epw.in/journal/2017/11/commentary/brics-and-new-financial-architecture.html?0=ip_login_no_cache%3Dfbaeb66444685ede82165f817e8e0053
- Tapscott, D., & Tapscott, A. (2016). Blockchain revolution: How the technology behind bitcoin is changing money, business, and the world. Penguin.
- United Nations Comtrade Database. (n.d.). Retrieved from https://comtrade.un.org/
- United Nations Economic and Social Commission for Asia and the Pacific, APTIAD (n.d.). Retrieved from https://artnet.unescap.org/aptiad/RCA.pdf
- World Bank. (2017). Global Value Chain Development Report 2017: Measuring and Analyzing the Impact of GVCs on Economic Development. World Bank Publications.22